\documentclass[twocolumn,apj,appendixfloats,numberedappendix]{emulateapj-rtx4}

\usepackage{natbib}
\usepackage{graphicx}
\usepackage{amsmath,amssymb}
\usepackage{graphicx}
\usepackage{txfonts}

\usepackage{ulem} 
\usepackage{natbib}
\newcommand{\vect}[1]{{\boldsymbol{#1}}}

\renewcommand{\d} {\mathrm{d}}
\newcommand{\pp}[2]  {\frac{\partial{#1}}{\partial{#2}}}

\newcommand{\bnabla}{\boldsymbol{\nabla}}

\begin{document}
\title {Anisotropy of third-order structure functions in MHD turbulence}
\author{Andrea~Verdini\altaffilmark{1}}
\email{verdini@arcetri.astro.it}
\affil{Dipartimento di fisica e astronomia, Universit\`a di Firenze,
Firenze}
\author{Roland~Grappin}
\affil{LPP, Ecole Polytechnique, Palaiseau, France.}
\author{Petr~Hellinger}
\affil{Astronomical Institute, AS CR, Prague, Czech Republic.}
\author{Simone~Landi}
\affil{Dipartimento di fisica e astronomia, Universit\`a di Firenze.}
\author{Wolf~Christian~M\"uller}
\affil{Technische Hochschule Berlin, Zentrum f\"ur Astronomie und
Astrophysik, Germany}
\altaffiltext{1}{Solar-Terrestrial Center of Excellence - SIDC, Royal
Observatory of Belgium, Bruxelles, Belgium.}

 \date{\today}
 
\begin{abstract}
The measure of the third-order structure function, $\vect{Y}$, is employed in
the solar wind to compute the cascade rate of turbulence. In the absence of a mean
field $B_0=0$, $\vect{Y}$ is expected to be isotropic (radial) and independent
of the direction of increments, so its measure yields directly the cascade rate. 
For turbulence with mean field, as in the solar wind, $\vect{Y}$ is expected to
become more two dimensional (2D), that is, to have larger perpendicular
components, losing the above simple symmetry. To get the cascade rate one should
compute the flux of $\vect{Y}$, which is not feasible with single-spacecraft
data, thus measurements rely on assumptions about the unknown symmetry.
We use direct numerical simulations (DNS) of magneto-hydrodynamic (MHD)
turbulence to characterize the anisotropy of $\vect{Y}$. We find that for
strong guide field $B_0=5$ the degree of two-dimensionalization depends on the
relative importance of shear and pseudo polarizations (the two components of an
Alfv\'en mode in incompressible MHD). The anisotropy also shows up in the
inertial range. The more $\vect{Y}$ is 2D, the more the inertial range extent
differs along parallel and perpendicular directions. We finally test the two
methods employed in observations and find that the so-obtained cascade rate may
depend on the angle between $B_0$ and the direction of increments. Both methods
yield a vanishing cascade rate along the parallel direction, contrary to
observations, suggesting a weaker anisotropy of solar wind turbulence compared
to our DNS. This could be due to a weaker mean field and/or to solar wind
expansion.
\end{abstract}
\keywords{The Sun, Solar wind, Magneto-hydrodynamics (MHD), Turbulence.}
   \maketitle
\section{Introduction.}
Magneto-hydrodynamic (MHD) turbulence in the
presence of a mean-field $B_0$ has a tendency to become two-dimensional (2D).
This tendency was early recognized by inspection of
the Fourier energy spectra in direct numerical simulations (DNS). The
energy distribution is indeed anisotropic, residing in
wavevector mostly perpendicular to the mean magnetic field 
\citep{Montgomery_Turner_1981,Shebalin_al_1983,Grappin_1986}.
Ideally one would like to quantify the
two-dimensionalization as a scaling relation between parallel and
perpendicular wavenumbers having the same energy density, $k_{||}\propto k_\bot^p$.
If $p=1$ the anisotropy is scale independent, and the aspect ratio
$k_{||}/k_\bot$ of the
isocontour of the Fourier spectrum does not change with scales. If instead
$p<1$ the aspect ratio increases with wavenumber, that is, the spectrum becomes more and more 2D at smaller and smaller scales. 
In DNS, the parallel spectral extent is generally very short, due the limited
achievable Reynolds numbers, and the parallel spectrum rarely shows a
power-law, rendering the distinction between scale-dependent and
scale-independent anisotropy difficult.
In contrast, the two-point correlation in real space expressed by
II-order structure functions, $S$, shows  in general nicer power-law scaling in
the parallel direction, allowing one to 
quantify the scale-by-scale anisotropy. In analogy with Fourier spectra, the
scaling relation involves parallel and perpendicular increments that have the same energy  $\ell_{||}\propto \ell_\bot^p$, also known as eddy anisotropy.

The II-order structure function anisotropy has been widely studied in DNS of
incompressible MHD turbulence. Using a local mean-field to identify parallel and perpendicular increments,  
one finds a scale-dependent anisotropy \citep{Cho_Vishniac_2000}:
the anisotropy grows at smaller and smaller scales, suggesting a complete 2D state at small enough scales. 
Furthermore, the anisotropy is controlled by $b_{rms}/B_0$ where $b_{rms}$ indicates the root-mean-square amplitude of turbulent fluctuations, the stronger $B_0$ the stronger the anisotropy \citep{Muller_al_2003}.
The anisotropy is also stronger in regions of stronger
magnetic field \citep{Milano_al_2001}.
Similarly, the anisotropy in the Fourier spectrum increases for the stronger $B_0$ \citep{Oughton_al_1998}.
However, employing structure function to measure the anisotropy with respect to
the global mean-field returns a scale-independent anisotropy \citep{Chen_al_2011}, implying that the two-dimensionalization does not increase at smaller scales but reaches an asymptotic value. 
A similar dichotomy exists in solar wind measurements, in which one can
compute the two-point correlation in time from time series of data
collected in-situ by spacecraft and then adopt the Taylor
hypothesis to obtain spatial increments. The increments are thus taken along
the radial direction, but the anisotropy with respect to the magnetic field
is recovered thanks to its variable direction with respect to the radial.
As in DNS, the structure function $S$ is found to be more energetic along perpendicular increments than along parallel increments. Again, a local mean-field analysis yields a scale-dependent anisotropy
\citep{Horbury_al_2008,Wicks_al_2010,Chen_al_2012}, while a
global mean-field analysis indicates a scale-independent anisotropy
\citep{Tessein_al_2009}. 
Several authors
\citep{Cho_al_2002,Chen_al_2011,Beresnyak_2012} showed that for strong
turbulence the scale-dependent anisotropy is smoothed out in a global mean-field
analysis, even in the presence of a strong mean-field. On the other hand, 
\citet{Matthaeus_al_2012} noted that since the local mean-field
is a random variable, the local II-order structure
functions involve higher-order statistics and cannot be the real space
equivalent of the power spectra. 
However for small enough $b_{rms}/B_0$ the global and local measures
are expected to coincide.

In this work we investigate the process of two-dimensionalization of MHD
turbulence, focusing on the anisotropy measured in the global frame.
In this frame, one can obtain a dynamical
equation (labelled KHYPP equation) that relates II-order and III-order structure functions \citep{PP}, extending to incompressible MHD the
Von Karman-Howart-Monin equation for incompressible hydrodynamic
turbulence. According to the KHYPP equation, for stationary and homogenous turbulence
in the inertial range, the divergence of the III-order structure function
$\vect{Y}$ is proportional to the cascade rate of turbulence
$4\epsilon=-\bnabla\cdot\vect{Y}$.
The divergence is negative, implying that the cascade is achieved by removing positive correlations, and thus increasing the amplitude of $S$
(flattening its power-law index).
Thus, by characterizing the anisotropy of $\vect{Y}$ one can get insight into
the process of two-dimensionalization of MHD turbulence.
Previous studies on the III-order structure functions in DNS of the
MHD equations were
limited to 2D \citep{Politano_al_1998, SorrisoValvo_al_2002} 
thus leaving out the issue of anisotropy.
In a recent work, \citet{Lamriben_al_2011} reported for the first time the
vector $\vect{Y}$ measured in an experiment of
rotating hydrodynamic turbulence. As rotation was increased, the 
anisotropy of the II-order structure functions also increased.
They found that the two-dimensionalization
can be associated with the tilting of the vector $\vect{Y}$ toward the plane
orthogonal to the rotation axis and that the tilting begins at small scales
and then propagates to larger and larger scales. 

In the present work, we 
carry out a similar analysis on data from three-dimensional (3D) DNS of
incompressible MHD turbulence by computing for the first time the 3D
III-order structure functions, in the presence or absence of a mean-field. 
We find that the degree of two-dimensionalization as measured by $S$
is associated to the relative excitation of pseudo and shear Alfv\'en
polarizations for stationary turbulence with mean field $B_0$. 
We also analyze the full KHYPP equation in developed decaying isotropic
turbulence, 
offering a description of the cascade in real space based on structure
functions (SF), analogous to the usual Kolmogorov cascade in Fourier space.
Note that
while the latter is based on the assumption of locality, the KHYPP
equation is free
from this assumption (requiring only homogeneity), thus representing a more
general description of the cascade process in MHD turbulence.

The III-order structure function $\vect{Y}$  has also been computed in solar wind data to obtain the cascade rate of solar wind turbulence \citep{SorrisoValvo_al_2007,Marino_al_2008,Marino_al_2012,Macbride_al_2008,Smith_al_2009,Stawarz_al_2009,Stawarz_al_2010}.
These rates are consistent with the heating rate estimated from proton temperature gradients
\citep{Vasquez_al_2007,Cranmer_al_2009}, suggesting that turbulence may
supply the heating required to sustain the non-adiabatic expansion of
the solar wind. 
However, applying the KHYPP equation to the solar wind is a bit problematic since solar wind turbulence is neither stationary nor homogenous (e.g. \citealt{Hellinger_al_2013,Gogoberidze_al_2013,Dong_al_2014}). 
To obtain the cascade rate one should compute the divergence of
$\vect{Y}$, which is quite difficult with
single-spacecraft data, since increments are taken only along one direction
at time (see however \citealt{Osman_al_2011} for an integral form that exploits
the four CLUSTER spacecraft with the minimal assumption of axisymmetry). The cascade rate can thus be retrieved only assuming a form for the unknown anisotropy of $\vect{Y}$. Although some theoretical predictions exist (e.g. \citealt{Podesta_al_2007,Galtier_al_2009}), two methods are commonly
employed in observations that assume respectively isotropy or an anisotropic
model based on the geometrical slab-plus-2D turbulence that was introduced by
\citet{Matthaeus_al_1990} to describe the two-point correlation function of solar wind turbulence. 
We will exploit the data from our DNS to test the two methods employed
in solar wind data against known anisotropic III-order structure functions and
estimate the possible systematic errors.

The plan of the paper is as follows.
In section \ref{sec2} we give a brief introduction to the KHYPP
equation, while in
section \ref{sec3} we describe the method
employed to compute structure functions (SF) of II-order and III-order. 
The results are presented in section \ref{sec4}, where we first describe the anisotropy of II-order structure functions of the
simulations considered. The rest of the section is dedicated to the III-order SF. We consider first a simulation of decaying turbulence without mean
magnetic field, allowing us to test the
soundness of our analysis method and to verify that the time-dependent
KHYPP equation
holds in developed decaying turbulence.
Then, we analyze two simulations of turbulence with mean-field that have
a different strength of anisotropy. 
Finally, in section \ref{sec5}, we test on runs with $B_0\ne0$ the two
methods employed to measure the cascade rate in the solar wind. 
We conclude with a discussion on the results and on the application to
the solar wind turbulence.

\section{Structure functions and the KHYPP equation.}\label{sec2}
The Von Karman-Howart-Yaglom, Politano-Pouquet equation (KHYPP) for non-stationary,
anisotropic, and incompressible MHD \citep{PP,Podesta_2008,Carbone_al_2009} can
be obtained from the original MHD equations written in
term of the Els\"asser variables $\vect{z^\pm}=\vect{u}\mp
\vect{b}/\sqrt{4\pi\rho}$, by subtracting the MHD equations evaluated at different
positions $\vect{x}$ and $\vect{x}+\vect{\ell}$ and by averaging in the volume. Under the assumptions of incompressible, homogeneous turbulence one obtains
\begin{eqnarray}
\partial_t \langle|\Delta\vect{ 
z^\pm}|^2\rangle+\bnabla_{\ell}\cdot
\left\langle\Delta \vect{z^\mp} |\Delta\vect{
z^\pm|^2}\right\rangle=\nonumber\\
-\Pi-\Lambda+2\nu\bnabla^2_\ell\langle |\Delta \vect{z^\pm}|^2\rangle
-4\epsilon^\pm,
\label{PP}
\end{eqnarray}
where we have defined the two-point correlation,
$\Delta\vect{ z^\pm}(\vect{ x},\vect{ \ell})=
\vect{ z^\pm}(\vect{ x}+\vect{ \ell})-\vect{ z^\pm}(\vect{ x})$, and $\langle...\rangle$ stands for the volume average.
These equations describe the evolution of the II-order structure function 
for each Els\"asser variable, $S^\pm=\langle|\Delta\vect{ z^\pm}|^2\rangle$.
The divergence term in the left hand side is the III-order structure function, 
$\vect{Y^\pm}=\left\langle
\Delta \vect{z^\mp}|\Delta \vect{z^\pm}|^2\right\rangle$, involving products of
$\Delta \vect{z^+}$, and $\Delta \vect{z^-}$, which we name Yaglom flux in
the following.
On the right hand side (rhs), $\Pi$ and $\Lambda$ represent pressure terms and sweeping terms (responsible for the Alfv\'en effect) respectively, both vanishing for globally homogeneous
turbulence \citep{Carbone_al_2009}. The remaining terms represent
dissipation. The first one involves the Laplacian with respect to the
increments $\ell$, it vanishes for vanishing viscosity (for simplicity we assumed equal viscosity and
resistivity $\nu=\eta$). The second one,
$\epsilon^\pm=-\partial_t E^\pm=\nu\langle\Sigma_j(\partial_j
z^\pm_i)(\partial_j z^\pm_i)\rangle$, is the dissipation rate of the
Els\"asser energies ($E^\pm=\langle|\vect{z^\pm}|^2/2\rangle$). In the
former, the derivatives of the primitive
fields $\vect{z^\pm}$ do not commutate with the averaging operation, and the
dissipation rate remains finite for vanishing viscosity.

Summing the contributions of both Els\"asser fields, one finally gets an expression for the total energy and cascade:
\begin{eqnarray}
\partial_t S+\bnabla_\ell\cdot \vect{Y}=-4\epsilon +2\nu\bnabla_\ell^2 S,
\label{PPT}
\end{eqnarray}
where $S=1/2(S^++S^-)$, $\vect{Y}=1/2(\vect{Y^+}+\vect{Y^-})$, and
$\epsilon=1/2(\epsilon^++\epsilon^-)$. Note that because of homogeneity in the
above Equations~\eqref{PP}-\eqref{PPT} all the variables depend only on the vector separation, $\vect{\ell}$. 

This equation is valid for decaying turbulence and describes the classical
scenario of a turbulent flow in which the dissipation of energy is
achieved through a cascade of energy toward smaller scales, where fluctuations
are finally damped by viscosity.
In forced turbulence one should add on the right hand side the forcing terms
(${\cal F}$) that inject energy (usually) at large scales. 

For stationary turbulence ($\partial_tS=0$) forced at large scales
(${\cal F}\ne0$ only at large scales) the
injection, cascade, and dissipation all occur at the same rate. 
At high Reynolds number one expects their respective ranges to be well
separated, in analogy to the Kolmogorov cascade in Fourier space, so one has that:
i) at large scales the second and last terms in Equation~\eqref{PPT} are
negligible and ${\cal F}=4\epsilon$, the forcing balances the
dissipation rate, ii) at small scales the second term is negligible, $2\nu\bnabla^2_\ell
S=4\epsilon$, and the damping rate is equal to the dissipation rate, 
and finally iii) at intermediate scales 
the last term is negligible, yielding
\begin{eqnarray}
\bnabla_\ell\cdot \vect{Y}=-4\epsilon, 
\label{PPI}
\end{eqnarray}
that is, the cascade rate, which is equal to the dissipation rate, is given by the divergence of the Yaglom flux.
Note that Equation~\eqref{PPI} can be used as a definition of the inertial
range as being the ensemble of scales for which the equation is approximately
satisfied. The definition should hold for quasi-stationary forced turbulence and
for developed decaying turbulence. As we will see, for developed
decaying turbulence, the time-dependent term is non-negligible only at large
scales where $\partial_t S=-4\epsilon$ (the dissipation is balanced by the decay
of the II-order SF at large scales).

We can now give a more physical interpretation of the cascade process by
rewriting Equation~\eqref{PPT} in terms of the
autocorrelation function, $C=C^++C^-$, with $C^\pm(\vect{\ell})=\langle
\vect{z^\pm}(\vect{x}+\vect{\ell})\cdot\vect{z^\pm}(\vect{x})\rangle$.
The autocorrelation functions are related to structure functions by
\begin{eqnarray}
S^\pm(\vect{ \ell})=2E^\pm-2C^\pm(\vect{ \ell}),
\label{SFAC}
\end{eqnarray}
and using $\partial_tE^\pm=-\epsilon^\pm$, the KHYPP equation becomes:
\begin{eqnarray}
\partial_t C-\bnabla_\ell\cdot \vect{Y}=-2\nu\bnabla_\ell^2 C,
\label{PPAC}
\end{eqnarray}
showing that $\vect{Y}$ is a flux of negative correlations. 
A permanent flux of negative correlations toward small scales is equivalent to
constantly building new small-scales gradients. For instance, 
the formation of 2D
quasi-perpendicular turbulence will be revealed by a Yaglom flux bringing
negative correlation at small perpendicular scales, hence the vector $\vect{Y}$ must be quasi-uniform, parallel to the $\ell_\bot$ axis, and pointing toward the parallel $\ell_{||}$ axis.

Coming back to Equation~\eqref{PPI}, for turbulence in stationary state or
decaying turbulence, the cascade is a constant at all inertial-range scales
($-\bnabla\cdot\vect{Y}=const$). Thus, the inertial range anisotropy cannot
appear as different cascade rates in the parallel and perpendicular
directions. The anisotropy instead will show up in the shape of the domain of
$\vect{\ell}$ for which $\bnabla\cdot\vect{Y}=const$. 

To illustrate such an anisotropy, one can assume some particular symmetry of the
flux $\vect{Y}$ to characterize the cascade, with the additional advantage of
obtaining a direct relation to the cascade rate, so as to avoid computing the divergence of $\vect{Y}$.
The simplest assumption is that of \textit{isotropic turbulence} for which
$\vect{Y}$ depends only on the scalar increment $\ell$. Rewriting
the divergence in spherical coordinates, and assuming stationary conditions
and vanishing viscosity, Equation~\eqref{PPI} in the inertial range becomes the
isotropic KHYPP equation, yielding 
\begin{eqnarray}
\epsilon^{iso}=-\frac{3}{4}\frac{Y_\ell(\ell)}{\ell},
\label{PPiso}
\end{eqnarray}
in which the Yaglom flux $Y_\ell=\vect{Y}\cdot\vect{\ell}/|\vect{\ell}|$ is
projected along the increment. This form is often used in solar wind studies,
since although one does not have access to the full divergence in in-situ data,
the cascade rate can be obtained directly from the projected Yaglom flux. 
The inertial range occupies a volume that is a
sector of a sphere, it can be defined as isotropic since it has the same extent and location on parallel and perpendicular increments. 

A strongly anisotropic case case is that of
\textit{2D turbulence}, obtained when the Yaglom flux in the inertial range depends only on the in-plane increments, $\vect{\ell}_\bot$:
\begin{eqnarray}
\bnabla\cdot\vect{Y}=\bnabla_\bot\cdot\vect{Y}_\bot=-4\epsilon ,
\label{PP2D}
\end{eqnarray}
where $\vect{Y}_\bot$ are the in-plane components of $\vect{Y}$ and
$\bnabla_\bot$ denotes derivatives with respect to the in-plane increments.
The Yaglom flux can have out-of-plane components, but the
cascade rate is determined only by the in-plane components. Note that 
a Yaglom flux having only in-plane components in the inertial range is a
sufficient condition to have a 2D cascade. Assuming isotropy of the in-plane
increments, that is, a dependence only on the scalar separation $\ell_\bot$, one obtains again a direct relation between III-order SF and the cascade rate:
\begin{eqnarray}
\epsilon^{2D}=-\frac{Y_{\ell_\bot}(\ell_\bot)}{2\ell_\bot},
\label{PP2DI}
\end{eqnarray}
with $Y_{\ell_\bot}=\vect{Y}\cdot\vect{\ell_\bot}/|\ell_\bot|$ indicating the
projection of $\vect{Y_\bot}$ on the radial direction in cylindrical coordinates.
Note that the inertial-range domain is not confined to the 2D plane even in
this anisotropic turbulence. As we will see, the anisotropy of the
inertial-range domain shows up in its different extent and location along
parallel and perpendicular increments.

For completeness, we consider finally the case of \textit{1D turbulence},
when the III-order SF depends only on one coordinate, say, $\ell_{||}$.
One obtains the cascade rate as
\begin{eqnarray}
\epsilon^{1D}=-\frac{Y_{||}(\ell_{||})}{4\ell_{||}}.
\label{PP1D}
\end{eqnarray}
Note that the geometrical model of \textit{slab turbulence} does not correspond
to the 1D turbulence, since in the former, fluctuations are assumed to
be perpendicular but to depend only on parallel wavevectors (and hence
parallel separations). The slab geometrical configuration would indeed have a
vanishing divergence. 

\begin{table}
\centering
\caption{Runs and parameters for simulations.}
\begin{tabular}{lcccccccl}
\hline
Run& $b_{rms}$& $B_0$& $R_x$& $\chi$& $N_x\cdot N_y\cdot N_z$& $10^{-4}\nu$&
$10^3~Re$ & forcing\\
\hline\hline
A & 0.5 & 0 & 1 & -   & $1024^3$          & $0.9$ & $2.6$ &decaying\\
  &     &   &   &     &                   &       & $\rightarrow2.2$ &  \\
B & 0.9 & 5 & 5 & 1.8 & $512^3$           & $1.5$ & $1.1$ &$k_{||}^f=1/5$\\
  &     &   &   &     &                   &       &        &$k_\bot^f\le2$\\
C & 1   & 5 & 1 & 0.2 & $256\cdot1024^2$ & $1$   & $2.8$ &$|k^f|\le2$       \\
  &     &   &   &     &                  &       &       &            frozen\\
\hline\hline
\end{tabular}
\label{table1}
\tablecomments{ $b_{rms}=\sqrt{2E_b}$ is the root mean square magnetic field
fluctuation. $B_0$ is
the mean magnetic field along the $x$ axis. $R_x=L_x/L_y$ is
the aspect ratio of the box of size $L_y=L_z=2\pi$.
The parameter
$\chi=t_A^f/t_{NL}^f= k_\bot^fb_{rms}/k_{||}^fB_0$ 
controls the strength of turbulence at forcing scales. $N_x,N_y,$~and~$N_z$ are the number of grid points. $\nu$ is the viscosity
coefficient (equal to the resistivity $\eta$). $Re=[2\pi/(k_\bot^fL_{diss})]^{4/3}$ is the effective Reynolds number, where the dissipation scale is
defined as $L_{diss}=(\nu^3/\epsilon)^{1/4}$. For run A, $Re$ decreases with time in the indicated interval. Forced wavenumbers are normalized by $L_y$, with $k_{||}=k_x$, $k_\bot=(k_y^2+k_z^2)^{1/2}$. In run B the sound speed is $C_S\approx 12$ and the conductivity coefficient is $\kappa=\nu$.}
\end{table}
\begin{figure*}[t]
\begin{center}
\includegraphics [width=0.98\linewidth]{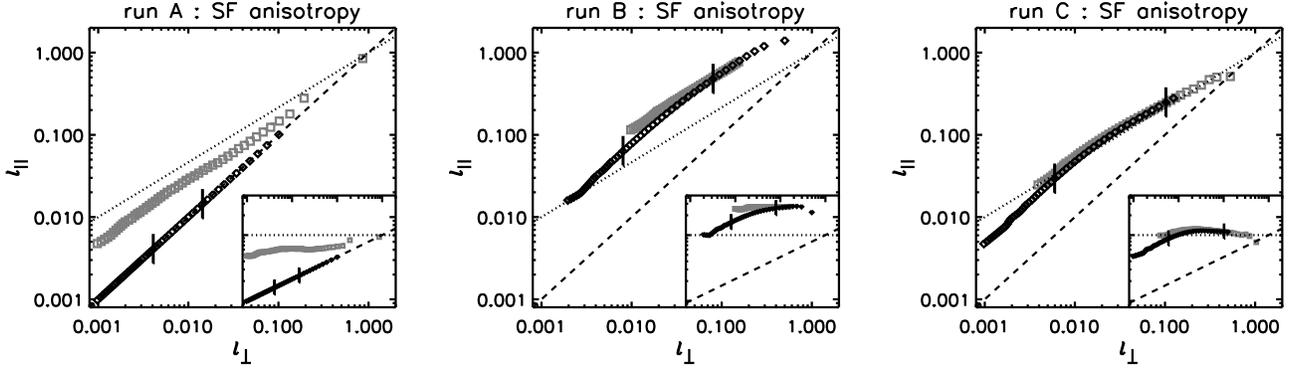}
\caption{Anisotropy of II-order structure functions
$\ell_\bot(\ell_{||})$ obtained by identifying the scales at which parallel and
perpendicular II-order SF have the same level, $S_{||}\equiv
S(\ell_{||},0)=S(0,\ell_\bot)\equiv S_\bot$.
Black diamonds and gray squares indicate the anisotropy 
with respect to the global or to the local mean-field respectively.
The vertical bars bound inertial range scales for the global SF as determined
by the III-order SF (see text). 
The two reference straight lines indicate scale-independent anisotropy
($\ell_{||}=\ell_\bot$, dashed line) and the critical balance scale-dependent
anisotropy ($\ell_{||}=\ell_\bot^{2/3}$, dotted line).
From left to right, run A (isotropic, decaying turbulence), run B (anisotropic,
forced, strong turbulence) and run C (anisotropic, forced, weak turbulence).
Insets display the same plots compensated by $\ell_\bot^{2/3}$.}
\label{fig1}
\end{center}
\end{figure*}
\section{Simulations and numerical method.}\label{sec3}
We consider three high-resolution simulations of MHD turbulence whose parameters are listed in table~\ref{table1}. 
Run A is a developed decaying simulation of
incompressible MHD turbulence without mean-field, representing
\textit{isotropic turbulence}. 
Run B is a simulation of weakly compressible MHD turbulence\footnote{The average
Mach number is $M_S=b_{rms}/C_s\approx 0.1$, where $C_S$ is the sound speed, 
while $M_S^{max}\approx 0.3$.}, with mean-field $B_0=5$ and anisotropic forcing. The forcing
is applied only to components perpendicular to the mean-field and to
wavevectors mainly perpendicular to the mean-field. Thus we are dealing
with \textit{strong anisotropic turbulence} of fluctuations with \textit{shear-Alfv\'en}
polarizations, a
configuration akin to Reduced MHD\footnote{A definition of shear and
pseudo polarizations is given in section~\ref{sec:pol}. For fluctuations with
mainly perpendicular wavevectors, pseudo polarizations have components parallel
to the mean field, while shear polarization have components perpendicular to
the mean field. In this limit, the former are absent in the Reduced MHD
formalism.}.
Finally run C is a simulation of incompressible MHD, again
with mean-field $B_0=5$, but with a forcing  which is isotropic in both
components and wavevectors. In the latter the forcing is actually a freezing of
the modes $1\le k\le2$ that maintains an equal amount of \textit{pseudo-Alfv\'en and
shear-Alfv\'en polarizations} at large scales, along with equipartition between magnetic
and kinetic energy and between Els\"asser energies.
This simulation can be classified as a case of \textit{weak
anisotropic turbulence} in terms of the strength parameter $\chi$ (see
table~\ref{table1}), although it does not have the properties
of classical weak turbulence
\citep{Ng_Bhatta_1997,Galtier_al_2000,Meyrand_al_2014}. Indeed, the 3D spectrum
has a relatively strong excitation in the parallel direction, resulting in a
peculiar anisotropy $E(k,\theta)=A(\theta)k^p$, with an isotropic spectral index
$p=-2-3/2$ in all directions (corresponding to a 1D spectrum with slope $-3/2$) and all the anisotropy appearing as a power
anisotropy at large scales $A(\theta)$. 
We will not discuss the properties and the origin
of such a spectrum that can be found in
\citet{Muller_Grappin_2005,Grappin_Muller_2010,Grappin_al_2013}); what is
mostly relevant for the present
analysis is that run C has a different 3D anisotropy compared to run B,
although in both runs energy resides mainly in perpendicular wavevectors.

We will use three measures to characterize the simulations: II-order structure
functions computed in the frame defined by the local mean-field (local $S$), II
and III order structure functions, respectively $S$ and $\vect{Y}$, computed in an absolute frame attached to the $x$-axis, which is the direction of the global mean-field when it is present. For all structure
functions, computation is made calculating increments in real space.
Local $S$ are obtained following method I of
\citep{Cho_Vishniac_2000}, i.e. the local field at scale $\vect{\ell}$ is
defined as
$\vect{B_0^\ell}(\vect{x})=1/2[\vect{b}(\vect{x}+\vect{\ell})+\vect{b}(\vect{x})]$.
Note that the measure of anisotropy, i.e. the ratio
$S(\ell_\bot,0)/S(0,\ell_{||})$ of local $S$, is not unambiguously defined.
In a turbulent medium fluctuations have a wide range of excited scales and the
definition of the mean field depends on both scale and position. Thus, higher-order statistics may be introduced in the local $S$ to a different extent,
depending on the averaging procedure.
However, the two-point average employed in this work is a good working
definition, at least in simulations of homogeneous turbulence since it was shown
to yield the same results as line average or volume averages, provided that the
averaging scale is smaller than the correlation length
\citep{Matthaeus_al_2012}.
The computation of local $S$ is made for the whole range of available increments $\vect{\ell}$ but the average is
made on a subset of grid points (typically $N_x\times N_y\times N_z=32^3$),
which was checked to be a sufficient statistics for the anisotropy to
converge.
On the other hand the III-order structure functions $\vect{Y}$ are signed
quantities and their computation requires large statistics to converge. Thus
given the number of grid points $N$ in a given direction, we compute $\vect{Y}$ either on a smaller range of increments ($\ell=(1...N/M)dx$ with typically $M=4$, $dx$ is the grid size), or on a coarser
grid of increments ($\ell=(1...N/M)Mdx$, with typically $M=4$), but
still on all the grid points, so the averages are made on a sample of $\gtrsim
10^7$ data. 
The dissipation rate
($\epsilon$) entering the KHYPP equation~\eqref{PPT}
is obtained directly from the 3D Fourier spectra of magnetic
and velocity field ($\hat{b}$ and $\hat{u}$ respectively),
\begin{eqnarray}
\epsilon=-\partial_t E=\nu\Sigma_k k^2\hat{u}^2_k+\eta\Sigma_k k^2\hat{b}^2_k.
\label{diss}
\end{eqnarray}

The terms appearing in Equation~\eqref{PPT} are evaluated for four
consecutive snapshots separated by approximately a nonlinear time (for the time 
derivative $\partial_t S$ we use a simple first-order scheme). All quantities
entering the KHYPP equation are normalized to the dissipation rate $\epsilon$ and then
averaged over the four snapshots. The 3D data are finally reduced for purposes
of representation and analysis by performing an isotropization (averaging over
polar and azimuthal angle $\theta$ and $\phi$ in spherical coordinates),
\begin{eqnarray}
\left.\bnabla\cdot\vect{Y}\right|_{iso}=\frac{1}{4\pi}\int\left(\pp{Y_x}{\ell_x}+\pp{Y_y}{\ell_y}+\pp{Y_z}{\ell_z}\right)\mathrm{sin}\theta\d\phi\d\theta,
\label{Yiso}
\end{eqnarray}
or an axisymmetrization (averaging over the azimuthal angle $\phi$ in
cylindrical coordinate with axis along the $\ell_x$),
\begin{eqnarray}
\left.\bnabla\cdot\vect{Y}\right|_{axis}=\frac{1}{2\pi}\int\left(\pp{Y_x}{\ell_x}+\pp{Y_y}{\ell_y}+\pp{Y_z}{\ell_z}\right)\d\phi.
\label{Yaxis}
\end{eqnarray}
In the following we drop the subscripts $iso$ and $axis$ and eventually mention explicitly the
average procedure used for representation. 
\section{Results}\label{sec4}
\subsection{Anisotropy of II-order structure functions}
We measured the anisotropy of II-order SF in two frames. In
the global frame the increments $\ell_{||}$ and $\ell_{\bot}$
are taken parallel and
perpendicular to a fixed direction $x$, which is the direction of the
mean-field $\vect{B_0}$ when it is present. In the local
frame the parallel and perpendicular directions are relative to
the scale-dependent mean-field direction $\vect{B_0^\ell}$ (see
section~\ref{sec2}). 
The measure of the anisotropy is obtained by identifying the couples of increments
($\ell_{||},~\ell_{\bot}$) at which the
parallel and the perpendicular SF have the same value:
$S_{||}\equiv S(\ell_{||},~0)=S_\bot\equiv S(0,~\ell_\bot)$, i.e. the
function $\ell_{||}(\ell_\bot)$ measures the 
aspect ratio of isolevels of the SF at different scales, also known as eddy shape. Scale-independent anisotropy results in a linear relation
$\ell_{||}\propto\ell_\bot$, that is, an aspect ratio $\ell_\bot/\ell_{||}$ that does not change with scale. 
Conversely, scale-dependent anisotropy results in a deviation from the
linear scaling $\ell_{||}\propto\ell_\bot^p$, with $p<1$: the aspect
ratio of SF, $\ell_\bot/\ell_{||}$ increases at smaller scales, that is, eddies
become more and more elongated in the parallel direction.
In figure~\ref{fig1} we plot $\ell_{||}(\ell_\bot)$ for the two measures of
anisotropy, local (gray squares) and global (black diamonds), for the three
runs listed in table~\ref{table1}. The dashed line, with slope $-1$, is a
reference for scale-independent anisotropy. 
The dotted line is a reference for the scale-dependent
anisotropy predicted by the critical balance relation \citep{GS95}:
\begin{eqnarray}
\ell_{||}\propto\ell_\bot^{2/3} .
\label{CB}
\end{eqnarray}
The insets display the same plots compensated by the critical balance 
anisotropy in order to better appreciate the scaling relation
$\ell_{||}(\ell_\bot)$.

In the local frame, all runs have a scale-dependent anisotropy extending to a
wide range of scales. The function $\ell_{||}(\ell_\bot)$ has a slope
flatter than $1$, thus the anisotropy grows with decreasing
scales and eddies are more and more elongated in the parallel direction. 
The scaling law actually follows the critical balance anisotropy Equation~\eqref{CB} 
in the range where $S_\bot\propto\ell_\bot^{2/3}$, extending for about a
decade for runs A, B, and C in the intervals
$0.008\lesssim\ell_\bot\lesssim0.05$, $0.02\lesssim\ell_\bot\lesssim0.2$,
and $0.008\lesssim\ell_\bot\lesssim0.08$ respectively. 
This critical balance anisotropy is quite robust since
in all runs the anisotropic relation holds rather well even outside the 
above-mentioned range of scales where $S_{||}$ and $S_\bot$ have a clear power-law
scaling.
Note that run C is a weak turbulence simulation, and it is not obvious that it
should have a critical balance anisotropy (see \citealt{Galtier_al_2005}
for an explanation based on a heuristic model of anisotropic turbulence). 

\begin{figure*}[t]
\begin{center}
\includegraphics [width=0.98\linewidth]{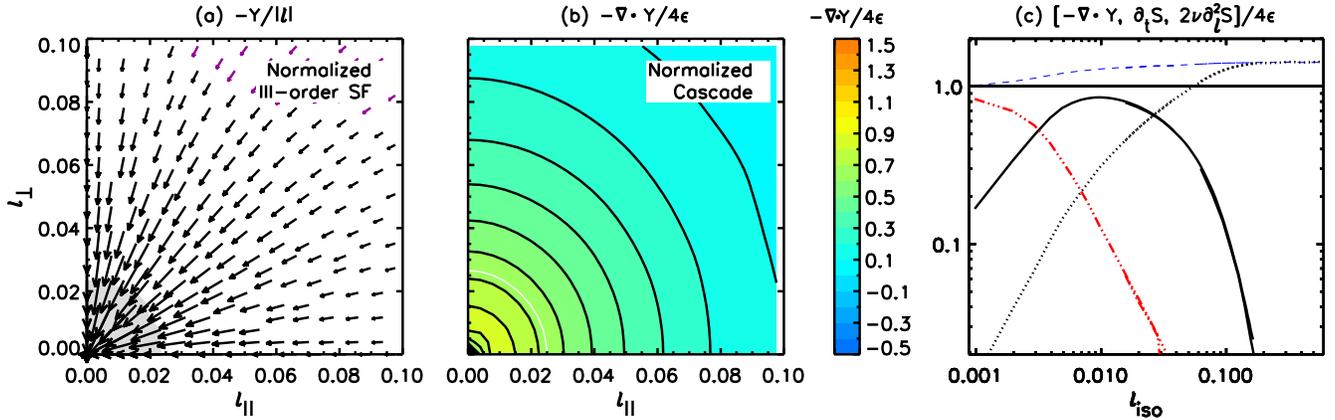}
\caption{Run A, $B_0=0$, isotropic decaying turbulence. \textit{Panel (a)}. 
III-order structure function, or Yaglom flux $\vect{Y}$, projected on the
($\ell_{||},~\ell_\bot$) plane and normalized by the scalar increment $\ell$.
Arrows are colored according to their angle $\theta_R$ with respect to the
radial direction (black is for $0^o\le\theta_R<5^o$, violet is for
$\theta_R\ge5^o$), their length is proportional to $|\vect{Y}|/\ell$.
\textit{Panel (b)}. Isocontour of $-\bnabla\cdot\vect{Y}$ normalized by the
dissipation rate $4\epsilon$. 
\textit{Panel (c)}. Comparison of different terms appearing in the KHYPP
equation,
Equation~\eqref{PPT}, after isotropization and normalization by $4\epsilon$: the divergence of the Yaglom flux $-\bnabla\cdot\vect{Y}$
(black solid line), the time dependent term $\partial_tS$ (dotted black line),
the dissipative term $2\nu\partial_l^2 S$ (red triple-dot-dashed line), and the sum
of the three terms (long-dashed blue line). The thick solid horizontal
line is a reference for $4\epsilon$. The gray area in panel (a) and the white
thick lines in panel (b) bound the scales at which 
$-\bnabla\cdot\vect{Y}$ is larger than the other terms
in the KHYPP equation, it is a rough estimate of inertial range scales.}
\label{fig2}
\end{center}
\end{figure*}
\begin{figure*}[t]
\begin{center}
\includegraphics [width=0.98\linewidth]{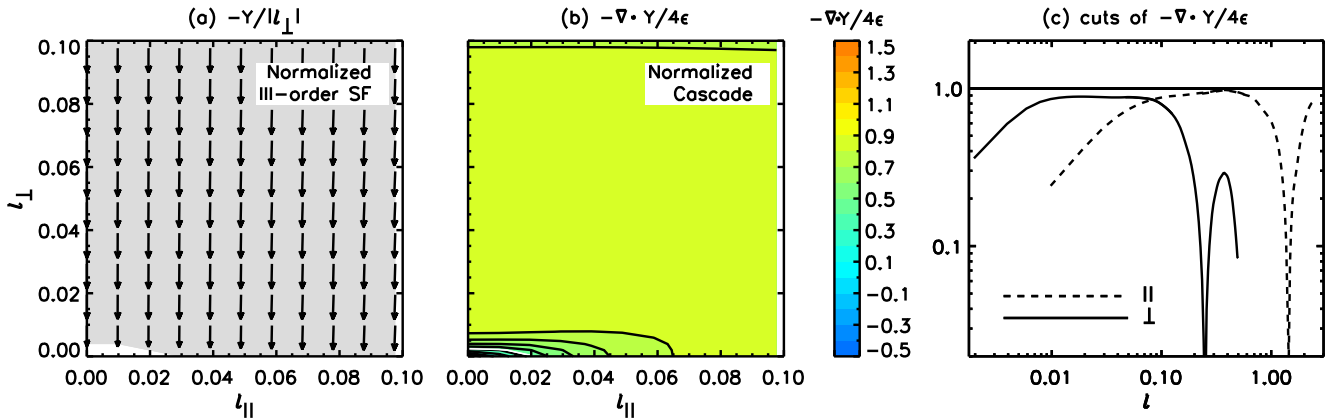}
\caption{Run B, $B_0=5$, strong turbulence.
\textit{Panel (a)}.
III-order structure function, or Yaglom flux $\vect{Y}$, projected on the
($\ell_{||},~\ell_\bot$) plane and normalized by the perpendicular increment
$\ell_\bot$.
Arrows are colored according to their angle $\theta_\bot$ with respect to the
perpendicular direction
direction (black is for $0^o\le\theta_\bot<5^o$),
their length is proportional to $|\vect{Y}|/\ell_\bot$.
\textit{Panel (b)}. Isocontour of $-\bnabla\cdot\vect{Y}$ normalized by the
dissipation rate $4\epsilon$. 
\textit{Panel (c)}. Cuts of $-\bnabla\cdot\vect{Y}$ in directions parallel (dashed line) and perpendicular (solid line) to
the mean-field $B_0$. The thick solid horizontal
line is a reference for $4\epsilon$.
The gray area in panel (a) and the white thick lines in panel (b) bound the scales at which $-\bnabla\cdot\vect{Y}$ is larger than the other terms
in the KHYPP equation, it is a rough estimate of inertial range scales.}
\label{fig3}
\end{center}
\end{figure*}

Consider now the anisotropy measured in the global frame (black diamonds in
Figure~\ref{fig1}),
which will be more relevant for the following analysis, since it is related to
the III-order SF by the KHYPP equation~\eqref{PPT}. As expected, run A is
perfectly isotropic, the aspect ratio is unity at all scales. 
The anisotropy of strong turbulence, run B, becomes scale-independent
(it has a slope equal to one) for scales $\ell_\bot\lesssim0.08$, approaching a constant ratio $\ell_{||}/\ell_\bot=A\approx10$ at small scales. For weak turbulence, run C, the anisotropy becomes scale-independent only at very small scales ($\ell_\bot\lesssim0.01$), with an aspect ratio
$A\approx5$ that is smaller compared to strong turbulence. 
The vertical bars in the plot bound the inertial range as identified
from Equation~\eqref{PPI}
\footnote{Following Equation~\eqref{PPI}, the inertial range is defined 
by the scales at which $|\bnabla\cdot\vect{Y}|$ is constant. 
We measure the slope of $\bnabla\cdot\vect{Y}$ in logarithmic scales
along the parallel and perpendicular directions and define inertial range
scales as those ones having a slope $\lesssim0.1$ (see Figure~\ref{fig3}c and Figure~\ref{fig4}c respectively). This is the procedure used for run B and C. 
We anticipate that for run A the divergence of the III-order SF is
not a nice straight horizontal line (Figure~\ref{fig2}c), so in this case the
inertial range is defined by the scales at which $|\bnabla\cdot\vect{Y}|$ is
about twice larger than all other terms in the KHYPP equation (corresponding to a slope
$\lesssim0.4$). As a partial cross-check, the chosen thresholds return a
dissipative scale that matches the scale at which $S_\bot/\ell_\bot$ has a
local maximum, which is the standard procedure for the identification of the
dissipative scale via II-oder SF.}.
While run B has a scale-independent anisotropy that develops in the inertial range, in run C scale-independent
anisotropy is attained only at dissipative scales.
Thus, at inertial-range scales the anisotropy is mostly scale dependent in our weak turbulence simulation (run C): it follows the critical balance scaling $\ell_{||}\propto\ell_\bot^{2/3}$ even when measured in the global frame, contrary to expectations \citep{Chen_al_2011}.
\subsection{III-order structure functions and KHYPP equation}\label{secIIIorder}
\subsubsection{Isotropic case}
We consider first the isotropic case (run A) for which $S$ is isotropic, so we
expect also to find an isotropic III-order SF. 
In Figure~\ref{fig3}, panel (a), we plot the Yaglom flux $\vect{Y}$ (III-order SF), averaged along the polar angle of cylindrical coordinates with axis
$\ell_x\equiv\ell_{||}$, and 
normalized by the scalar increment $\ell$. We consider relatively small scales
(the largest scale is $\ell=0.5$) to highlight inertial range features, as will
be clearer below. The Yaglom flux is almost
radial at large scales and becomes remarkably radial at smaller scales
$\ell\lesssim0.08$. The length of the arrows increases
toward the origin, indicating that the intensity of the cascade increases
when approaching the inertial range (the gray area) while keeping the same
(radial) direction. Note also that the arrow length is constant on circles of
given $\ell$, meaning that $\vect{Y}\approx \mathrm{Const}\times\vect{\ell}$ as
expected for isotropic turbulence, Equation~\eqref{PPiso}.

In panel (b) we plot the isocontours of the divergence of the Yaglom flux,
normalized by the dissipation rate. The isocontours are roughly isotropic at large scales and become perfectly
isotropic at small scales ($\ell\le0.03$). In a small interval of scales around 
$\ell\approx0.01$,  the divergence $-\bnabla\cdot\vect{Y}$ has a maximum
reaching the value $\approx0.9$. Thus, the dissipation rate is approximately
equal to the cascade rate, and these scales can be identified as the
inertial range of turbulence. However, the divergence is not strictly a
constant, as expected for the inertial range. Note that although the latter is
very short, it is uniformly distributed among scales.

Finally, in panel (c) we plot in logarithmic scales, after isotropization and normalization
by $4\epsilon$, all the terms appearing in the KHYPP equation, Equation~\eqref{PPT}, namely,
the divergence of the Yaglom flux $-\bnabla\cdot\vect{Y}$
(thick solid line), the dissipative term $2\nu\partial_\ell^2 S$ (red 
triple-dot-dashed line), and the time-dependent term $\partial_t S$ (dotted
line).
The dissipative term dominates at small scales, while the time-dependent term
(decay) dominates at large scales.
The cascade term, $-\bnabla\cdot\vect{Y}$, is larger than the other terms
for scales $0.003\lesssim\ell\lesssim0.03$ (the gray area in panel (a)). These
two extrema can be identified with the injection scale and the dissipative
scale,
respectively. It is worth noting that the dissipation scale defined in this way coincides with the estimate based on II-order SF in Figure~\ref{fig1}. 
From this logarithmic plot it is clear that the inertial range is quite small
in this decaying simulation because $-\bnabla\cdot\vect{Y}$ is much
larger than other terms only in a small interval centered at $\ell\approx0.01$, where it is roughly horizontal and equal to 0.9
(it should be equal to one in an ideally infinite inertial range). 

In the same plot we also traced, as a blue long-dashed line, the sum of the
three terms just discussed, that should amount at all scales to the dissipation
rate $4\epsilon$ (thick solid horizontal line) for good energy conservation.
The sum is an almost horizontal straight line, only a factor $1.2$ higher
than the dissipation rate for scales
$\ell\gtrsim0.005$. This confirms that the statistics is large enough to ensure convergence and that the
conservation of energy holds with sufficient accuracy except at very small
scales where a small numerical dissipation probably kicks in: the injection at large scale (including the decay), the cascade in the inertial range, and the
dissipation at small scales all occur at the same rate.

To summarize, the analysis of the isotropic turbulence confirms the
theoretical expectations: inside the
inertial range the Yaglom flux is radially directed and its magnitude scales linearly with the scalar increment
$\ell$. The divergence of the Yaglom flux is approximately constant in
the inertial range, and it is uniformly distributed among scales (isotropic). 
Note, however, that the extent of the inertial range is very limited due to the
relatively small Reynolds number that prevents the formation of a large range
of scales where the divergence of the Yaglom flux is the dominant term. With
the current resolution ($1024^3$) it is at best half a decade,
indicating that this is the minimal resolution required for studies of decaying
turbulence (although hyperviscosity would probably alleviate the problem).
This is relevant for solar wind studies \citep{Dong_al_2014},
in which expansion induces an additional decay of magnetic and kinetic energy
on top of the decay due to the turbulent dissipation. 

\begin{figure*}[t]
\begin{center}
\includegraphics [width=0.98\linewidth]{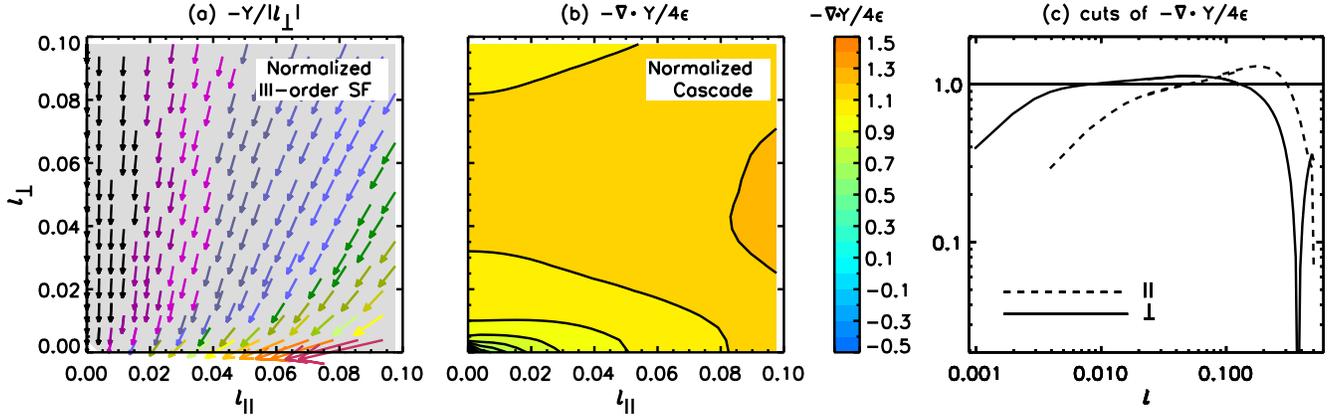}
\caption{Run C, $B_0=5$, weak turbulence.
Same as in Figure~\ref{fig3}. In panel (a) the arrow colors,
from black to purple, indicate $\theta_\bot\in[0^o,90^o]$ binned
in intervals of $5^o$.}
\label{fig4}
\end{center}
\end{figure*}
\subsubsection{Anisotropic case, strong turbulence}
Let's now turn to the anisotropic run B, which is a simulation of strong
turbulence with guide field and forced at large scales on components perpendicular $B_0$.
In Figure~\ref{fig3}, panel (a) we plot again the III-order SF, i.e. the 
Yaglom flux $\vect{Y}$, averaged over the polar angle in cylindrical coordinates
with axis along increments parallel to the mean magnetic field. At variance
with the isotropic case, the arrows are colored according to the angle
$\theta_\bot$ formed with respect to the perpendicular direction (black
color stands for 
$\theta_\bot\le5^o$), and their length is normalized to the perpendicular
increment $\ell_\bot=(\ell_y^2+\ell_z^2)^{1/2}$. The Yaglom flux is remarkably
vertical in the inertial range (the gray area), and it is proportional to the
perpendicular increments $\vect{Y}\propto\vect{\ell_\bot}$ (the arrow length
is uniform in the whole inertial range, after normalization). This suggest that
turbulence is undergoing a purely 2D cascade, Equation~\eqref{PP2DI}.

In panel (b) the normalized divergence of the Yaglom flux is constant over a
large interval of parallel and perpendicular increments, with a value close to
1 (the light-green area at level $0.9$). However it is not uniformly distributed among scales, the isocontour of
constant divergence extends to smaller perpendicular scales, suggesting
that the cascade is not removing positive correlation from the parallel
direction. This is consistent with
Figure~\ref{fig1}b that shows a clear anisotropy of $S$ in favor
of a two-dimensionalization in the perpendicular plane.

This can be better appreciated in panel (c), where we plot cuts along the
parallel and perpendicular directions of the divergence of the Yaglom flux in
logarithmic scales. There is a very nice constant divergence in the perpendicular direction, with a value
close to the dissipation rate (the horizontal thick solid line at level 1),
covering about one decade in the range $0.01\lesssim\ell_\bot\lesssim0.1$. In the parallel direction, an approximate inertial range is also found, having a
smaller extent and shifted to larger scales $0.1\lesssim\ell_{||}\lesssim0.6$.

We recall that from 
$-\bnabla\cdot\vect{Y}$ (panels (b) and (c))
one can identify the location of the inertial range and its
distribution among parallel and perpendicular scales. On the other hand, the
simple dependence of the Yaglom flux allows us to identify the cascade as a 2D
cascade with $\vect{Y}=-2\epsilon\vect{\ell_\bot}$
We anticipate that 
$\vect{Y^\mp}=\left\langle\Delta \vect{z^\pm}|\Delta \vect{z^\mp}|^2\right\rangle$ have only perpendicular components because
there is a dominance of shear-Alfv\'en polarizations. Indeed, these polarizations have $\Delta
\vect{z^\pm}$ lying in the perpendicular plane and, as a consequence, the cascade
is 2D. This simple picture of 2D cascade does not hold anymore as soon as
pseudo polarizations, which have out-of-plane components, are energetically important.

\subsubsection{Anisotropic case, weak turbulence}
We finally consider the case of weak turbulence with guide field (run C), which is forced isotropically at small wavevectors $1\le|\vect{k}|\le2$ by imposing at all times (freezing) the
corresponding Fourier modes of the fluctuating fields $\vect{B},~\vect{u}$
(note that their $x,y,z$ components have also equal energy).
In fig.\ref{fig4}, panel (a) one can see immediately that the Yaglom flux is
oblique, with an angle $\theta_\bot$ that changes with scales (we normalize
arrow length by the perpendicular increment $|\vect{Y}|/\ell_\bot$
as in Figure~\ref{fig3}). This means that
the III-order SF has a parallel component and a non-negligible dependence on
the parallel increments, thus contributing to the cascade rate through the
divergence of the Yaglom flux. Such contribution seems to decrease at small
parallel scales, where the Yaglom flux becomes vertical, hinting to a milder
two-dimensionalization of this weak turbulence cascade.

Isocontours of $-\bnabla\cdot\vect{Y}$ are plotted in panel (b), with the usual
normalization by $4\epsilon$. The inertial range can be identified with the
light-orange area at level $1.1$, extending to a wide range of parallel and
perpendicular scales. Its distribution is non-uniform and more complex than
that of strong turbulence (Figure~\ref{fig3}b), reflecting the dependence of
the III-order SF from $\ell_\bot$ and $\ell_{||}$.

Panel (c) shows cuts of the divergence of the Yaglom flux along the parallel and perpendicular directions. Although the
divergence is not exactly constant, the inertial range
covers more than one decade in the
perpendicular direction, $0.005\lesssim\ell_\bot\lesssim0.1$, yielding a cascade rate that is slightly higher than
the dissipation rate $4\epsilon$. On the other hand, the cut in the parallel
direction is less flat, making more questionable the identification of
an inertial range in this direction. The approximate parallel inertial range is
shorter and located at larger scales,
$0.03\lesssim\ell_{||}\lesssim0.3$. 

\subsubsection{The KHYPP equation for pseudo and shear Alfv\'en
polarizations}\label{sec:pol}
Decomposing fluctuations in pseudo and shear Alfv\'en
polarizations proves useful to analyze in some more detail the relation between
the Yaglom flux and the cascade rate in run C, which is a simulation of
incompressible MHD. 
The decomposition is made in Fourier space, where pseudo Alfv\'en polarizations and shear Alfv\'en polarizations are oriented along the unitary vectors (e.g.
\citealt{Maron_Goldreich_2001}),
\begin{eqnarray}
\vect{\xi_{sh}}= \frac{\vect{k}\times\vect{B_0}}
{[1-(\vect{k}\cdot\vect{B_0})^2]^{1/2}},\;\;\;\;
\vect{\xi_{ps}}= \frac{\vect{B_0}-(\vect{k}\cdot\vect{B_0})\vect{k}}
{[1-(\vect{k}\cdot\vect{B_0})^2]^{1/2}}.
\label{eq:vers}
\end{eqnarray}
This decomposition is completely equivalent to the decomposition into
toroidal and poloidal components of the magnetic fluctuations. 
In incompressible MHD the same decomposition applies to the velocity field,
which is also solenoidal.
Shear Alfv\'en polarizations are the proper Alfv\'en
modes in full MHD. Their component is perpendicular to both the mean field $\vect{B_0}$ and the wavevector $\vect{k}$, being incompressible. The pseudo Alfv\'en polarizations are the incompressible limit of slow modes in MHD.
Their component lies in the plane identified by the mean field and the
wavevector, and it is again perpendicular to the wavevector.  For fluctuations
with strong anisotropic spectra ($k_\bot>>k_{||}$), shear
polarizations have wavevectors and components lying in the plane
perpendicular to $B_0$, thus they represent the 2D modes in the slab-plus-2D
decomposition introduced by \citet{Matthaeus_al_1990}. Instead, pseudo polarizations have wavevectors in the
plane perpendicular to $B_0$ but components along $B_0$. This polarization,
which is absent in Reduced MHD, is instead present in 2.5D configurations with
out-of-plane mean field, and should not be confused with the slab
component that has wavevectors parallel to $B_0$ and fluctuations perpendicular
to it (and is thus included in Reduced MHD). 

After decomposing fluctuations in pseudo and shear Alfv\'en polarizations, we go back to the real space and compute separately
all the contributions to the KHYPP equation, Equation~\eqref{PPT}. 
Note that for strictly parallel wavevectors 
the pseudo-shear decomposition degenerates, so we remove such modes (the
slab component) in the following analysis to avoid arbitrary partition of energy into the shear and pseudo polarizations.
Using Parseval's theorem
$\langle\Delta \vect{z^\pm_{ps}}\cdot\Delta \vect{z^\pm_{sh}}\rangle=0$, the
decomposed KHYPP equation can be written as:
\begin{eqnarray}
&\partial_t& (S_{sh}+ S_{ps})
-2\nu\bnabla^2_\ell (S_{sh}+S_{ps})+4\epsilon=\label{yagdec}\\
&-&\bnabla_{\ell}\cdot \left\langle
\Delta\vect{z_{sh}}|\Delta\vect{z_{sh}}|^2 +
\Delta\vect{z_{sh}}|\Delta\vect{z_{ps}}|^2\right\rangle \label{yags}\\
&-&\bnabla_{\ell}\cdot \left\langle
\Delta\vect{z_{ps}}|\Delta\vect{z_{sh}}|^2 +
\Delta\vect{z_{ps}}|\Delta\vect{z_{ps}}|^2\right\rangle \label{yagp}\\
&-&\bnabla_{\ell}\cdot \left\langle
2\Delta\vect{z_{sh}}(\Delta\vect{z_{sh}}\cdot\Delta\vect{z_{ps}})+
2\Delta\vect{z_{ps}}(\Delta\vect{z_{sh}}\cdot\Delta\vect{z_{ps}})\right\rangle,
\label{yagps}
\end{eqnarray}
where we summed the $\pm$ species and dropped $\pm$ superscripts,
i.e. $S_{sh}=1/2\langle|\Delta
\vect{z^+_{sh}}|^2+|\Delta\vect{z^-_{sh}}|^2\rangle$,
 $\Delta \vect{z_{sh}}|\Delta \vect{z_{ps}}|^2=1/2[\Delta
\vect{z^+_{sh}}|\Delta\vect{z^-_{ps}}|^2+\Delta
\vect{z^-_{sh}}|\Delta\vect{z^+_{ps}}|^2]$, etc ...

The III-order SF on the rhs is split into three lines containing respectively
(1) the strain of the shear polarizations on the shear and pseudo energies, (2) the
strain of the pseudo polarizations on the shear and pseudo energies, and (3)
the mixed terms accounting for the exchange of energy during the cascade
between the shear and pseudo polarizations. We anticipate that the mixed terms
are negligible in the inertial range, thus 
we split the above equation into a system of equations for $S_{ps}$ and
$S_{sh}$, with their own cascade rate $\epsilon_{sh}$ and $\epsilon_{ps}$,
\begin{eqnarray}
\partial_t S_{sh}
-2\nu\bnabla^2_\ell S_{sh}+4\epsilon_{sh} =
-\bnabla_{\ell}\cdot \left\langle
\Delta\vect{z_{sh}} |\Delta\vect{z_{sh}}|^2 +
\Delta\vect{z_{ps}} |\Delta\vect{z_{sh}}|^2 
\right\rangle,\nonumber\\
\label{SSshps}
\\
\partial_t S_{ps}
-2\nu\bnabla^2_\ell S_{ps}+4\epsilon_{ps} =
-\bnabla_{\ell}\cdot \left\langle 
\Delta\vect{z_{sh}} |\Delta\vect{z_{ps}}|^2 +
\Delta\vect{z_{ps}} |\Delta\vect{z_{ps}}|^2 
\right\rangle.\nonumber\\
\label{PPshps}
\end{eqnarray}
Note that each equation contains only quadratic terms of a given Alfv\'en
polarization (shear or pseudo) even in the III-order SF. On the rhs of
Equations~\eqref{SSshps}-\eqref{PPshps}, the second terms represent an active
contribution to the cascade of pseudo Alfv\'en polarizations: if it is
vanishing or negligible the pseudo Alfv\'en polarizations can be said passive.

\begin{figure*}[t]
\begin{center}
\includegraphics [width=0.98\linewidth]{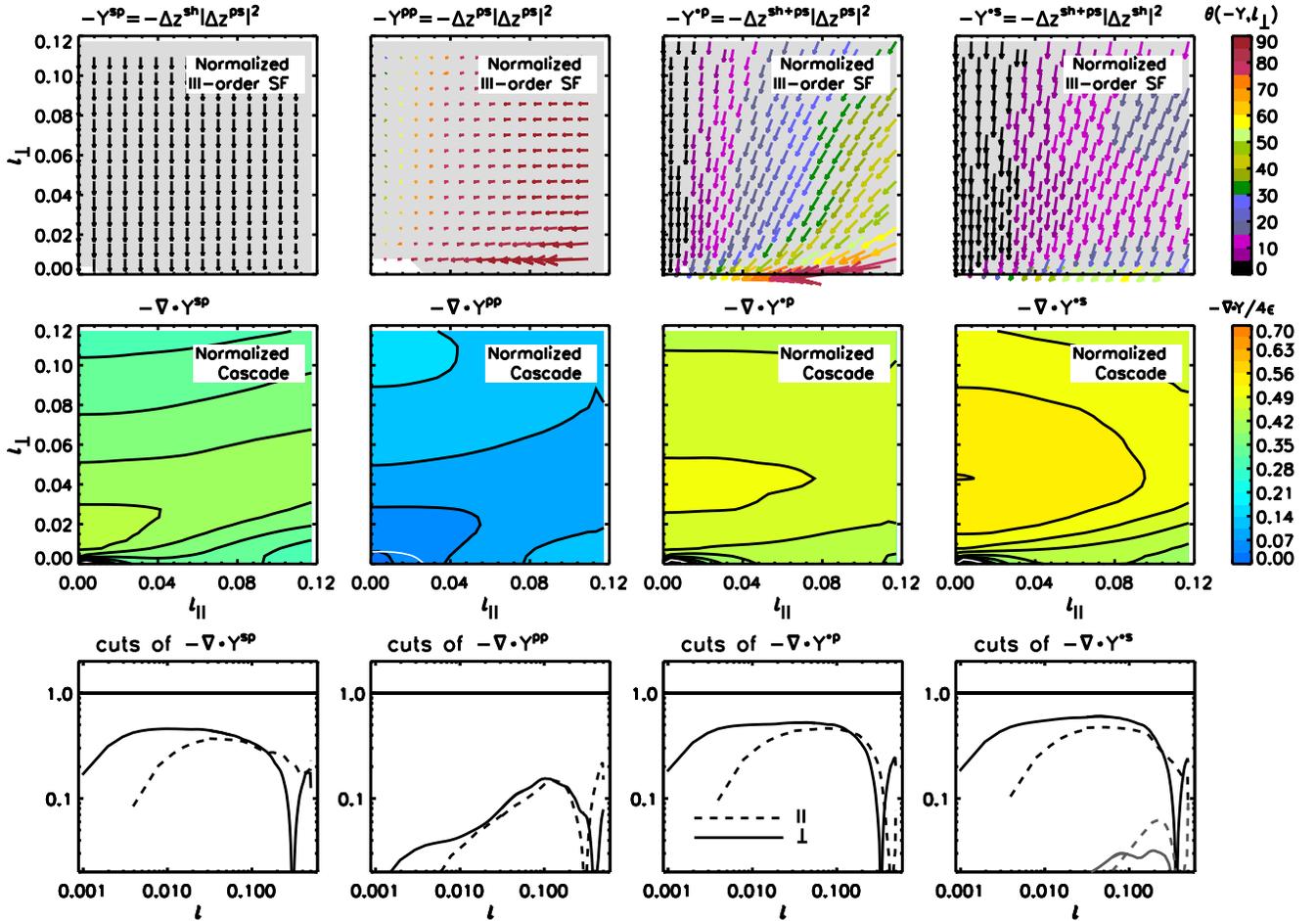}
\caption{Run C. Decomposition of the KHYPP equation for shear and pseudo
energies. As in figs.~\ref{fig3},~\ref{fig4}, $\vect{Y}$ is
normalized by $\ell_\bot$, $-\bnabla\cdot\vect{Y}$ is normalized
by $4\epsilon$. 
In the first three columns we analyze the KHYPP equation for pseudo
energy, Equation~\eqref{PPshps}, by plotting the
contribution to the cascade appearing on the rhs of Equation~\eqref{PPshps}. The first
term, i.e. the strain of shear polarizations on
the pseudo energy (column 1); the second term, i.e. the strain of
pseudo polarizations on pseudo energy (column 2); and their sum (column 3).
In column 4 we consider the cascade for shear energy, Equation~\eqref{SSshps}, without
separating the pseudo and shear contribution. In the bottom panel, fourth
column, we also plot with grey color
the contribution of the mixed terms, Equation~\eqref{yagps}.}
\label{fig5}
\end{center}
\end{figure*}
This separation will appear justified by the analysis of each term, which
is presented in Figure~\ref{fig5} in the same format of
figs.~\ref{fig3},~\ref{fig4}, i.e. with the the Yaglom flux normalized by the
perpendicular increment, $\vect{Y}/\ell_\bot$, and the divergence normalized by
the cascade rate, $-\bnabla\cdot\vect{Y}/4\epsilon$.
We consider separately the two contributions in the rhs terms of the
equation for the pseudo energies, Equation~\eqref{PPshps}.
In the first column we have the III-order SF
accounting for the strain of shear polarizations acting on the pseudo
energy, $\vect{Y}^\mathrm{SP}=\left\langle 
\Delta\vect{z_{sh}} |\Delta\vect{z_{ps}}|^2\right\rangle$ (the first term in
the rhs). The Yaglom flux (upper panel) is perpendicular, as in run B, since
the vectorial part of $\vect{Y}^\mathrm{SP}$ is made of only shear
polarizations, which have only perpendicular
components. Its magnitude depends mostly on $\ell_\bot$ as can
be guessed by the length of the arrows, with a small dependence on
$\ell_{||}$ at large parallel scales. 
This dependence is better appreciated on the map of
$-\bnabla\cdot\vect{Y}^\mathrm{SP}$ (middle panel) that has
slightly inclined isocontours instead of horizontal isocontours. The inertial
range identified with the green area at level $\approx0.4$ is non uniformly
distributed in the $(\ell_{||},~\ell_\bot)$ plane, occupying preferentially perpendicular scales
$\ell_\bot<0.05$ independently of parallel scale.
In the bottom panel one can better locate the inertial range which is
much more extended toward smaller scales in the perpendicular than in
the parallel direction. 

In the second column, we analyze the III-order SF accounting for the strain of
pseudo polarizations on the pseudo energy, $\vect{Y}^\mathrm{PP}=\left\langle 
\Delta\vect{z_{ps}} |\Delta\vect{z_{ps}}|^2\right\rangle$, the second term in the rhs of Equation~\eqref{PPshps}. 
Note that the Yaglom flux is now horizontal (excepts at
small parallel scales $\ell_{||}\lesssim0.03$), since for a spectrum (or
II-order SF) with energy residing mainly in perpendicular wavevectors,
pseudo polarizations have a dominant parallel component.
The contribution to the divergence (middle panel) is almost
complementary to the previous term, being larger at large parallel and
perpendicular scales while it is negligible in the inertial range. This is
better seen in the bottom panel, in which it is also shown clearly
that this III-order SF does not form 
an inertial range on its own, but it is responsible for a non-constant energy flux from large to small scales in both parallel and perpendicular directions. 
One would be tempted to model the two terms, $\vect{Y}^\mathrm{SP}$ and
$\vect{Y}^\mathrm{PP}$, as 2D+1D components respectively, according to the orientation of the Yaglom flux.
However, such a decomposition does not work since the candidate for the 1D
component ($\vect{Y}^\mathrm{PP}$) has a
strong dependence on $\ell_\bot$ (second column, top panel),
and its divergence does not yield any identifiable inertial range (second column, middle and bottom panels).

Finally in the third column we analyze the sum of the two terms just discussed,
$\vect{Y}^\mathrm{*P}=\vect{Y}^\mathrm{PP}+\vect{Y}^\mathrm{SP}=\left\langle(\Delta\vect{z_{sh}}+\Delta\vect{z_{ps}})|\Delta\vect{z_{ps}}|^2\right\rangle$, the whole rhs of Equation~\eqref{PPshps}. The Yaglom flux is now oblique, with
dependence on both $\ell_{\bot}$ and $\ell_{||}$, the latter being mostly
inherited from the strain of pseudo polarizations ($\vect{Y}^\mathrm{PP}$),
indicating a smaller degree of two-dimensionalization for turbulence of pseudo
Alfv\'en polarizations, compared to the strong turbulence of run B. 
In the middle panel the inertial range extends to parallel and
perpendicular scales in a way somewhat similar to run B, although
the Yaglom flux is quite different.
Comparing the bottom panels in columns one, two, and three, one can see that 
the strain of both shear and pseudo polarizations contribute to the formation of the wide
inertial range in the perpendicular direction
$0.005\lesssim\ell_\bot\lesssim0.1$, as well as to the shorter inertial range
found at larger scales in the parallel direction
$0.03\lesssim\ell_\bot\lesssim0.2$ (their ranges coincide with those one
obtained without separating shear and pseudo polarizations).
More precisely, shear polarizations contribute with a constant cascade at small scales
($\ell\lesssim0.04$), while
pseudo polarizations control the injection of energy into the cascade at large scales ($\ell\approx0.1$).

We do not repeat the analysis of the strain of shear and pseudo polarizations
on shear energy, Equation~\eqref{SSshps}, since it yields qualitatively similar
results. The only noticeable difference is that the Yaglom flux
$\vect{Y}^\mathrm{PS}=\left\langle\Delta \vect{z^{ps}}|\Delta\vect{
z^{sh}}|^2\right\rangle$ is radially directed 
instead of being horizontal. We
plot in the fourth column the whole contribution to the cascade of shear
energy,
$\vect{Y}^\mathrm{*S}=\left\langle(\Delta\vect{z_{sh}}+\Delta\vect{z_{ps}})|\Delta\vect{z_{sh}}|^2\right\rangle$.
Comparing the third and fourth columns, one can see that
(i) shear and pseudo energy cascade qualitatively in a similar way (top
panels), although the Yaglom flux for shear polarizations,
$\vect{Y}^\mathrm{*S}$, is more perpendicular (more 2D);
(ii) the cascade rate is slightly stronger for shear energy than for
pseudo energy (middle panels); 
and (iii) a clear inertial range is
found for both shear and pseudo energies (bottom panel) following the
decomposition of Equations~\eqref{SSshps}-\eqref{PPshps}. 

In the fourth column, bottom panel, we also plot the sum of the mixing terms,
Equation~\eqref{yagps}, along the parallel and perpendicular increments (gray lines). 
They are negligible compared to all other terms in the inertial range,
indicating that the shear and pseudo polarizations cascade without exchanging
their energy, as already pointed out by \citet{Maron_Goldreich_2001} for
decaying simulations. However, at large scales $\ell\approx 0.2$, they are of the same order of the term accounting for the strain exerted by pseudo polarizations (II column, bottom panel in
Figure~\ref{fig5}), suggesting an exchange of energy
between the shear and pseudo polarizations.

To summarize, in run C the freezing of large scales (forcing) causes an
exchange of energy between pseudo and shear polarization at large scales. 
The two polarizations cascade in a similar manner (same rate, same
inertial range) under the action (strain) of both the shear and pseudo
polarizations. 
The flux toward small scales
that is triggered by pseudo polarizations is smaller in magnitude and not
constant (it is not a proper inertial range on its own), but it affects considerably the total cascade rate at
large inertial-range scales ($\ell\approx0.1$).
On the other hand, the shear polarizations control the constant energy flux at
small inertial-range scales ($\ell\lesssim0.04$). 
Thus according to the separation made in
Equations~\eqref{SSshps}-\eqref{PPshps},
the two cascades are not totally independent. Indeed, whatever the
polarization considered ($S_{ps}$ or $S_{sh}$), the cascade is triggered by
pseudo polarizations at the interface between injection scales and
inertial range scales, while it is sustained by shear polarizations
at smaller inertial range scales. 
The presence and importance of the pseudo polarizations is revealed by the Yaglom flux, the more pseudo polarizations are energetic, the more $\vect{Y}$ becomes oblique.
\begin{figure*}[t]
\begin{center}
\includegraphics [width=0.98\linewidth]{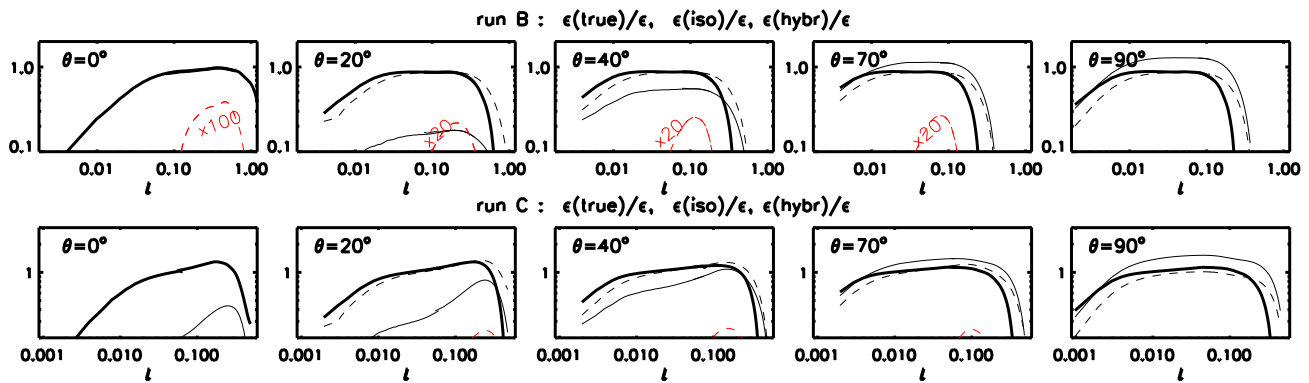}
\caption{Comparison of the true cascade rate, Equation~\eqref{ctrue} (solid thick line)
with the isotropic cascade rate, Equation~\eqref{ciso} (thin solid line), and the hybrid
cascade rate, Equation~\eqref{caniso} (dashed black line). The red dashed line is the
1D (parallel) cascade rate obtained within the hybrid method.
From left to right, panels refer to increasing angles $\theta$ between the sampling direction $\vect{\ell}$ and the mean-field $\vect{B_0}$.} 
\label{fig6}
\end{center}
\end{figure*}
\section{Anisotropy of the III-order SF and solar wind measurements}\label{sec5}
In measuring the cascade rate of solar wind turbulence through III-order SF,
one needs to make assumptions on the unknown anisotropy of $\vect{Y}$. Despite
the limitations due to the imposed symmetries, this
brings the advantage of increasing the statistic, on the one hand, and of avoiding the computation of the divergence of $\vect{Y}$ on the other hand.
Two common assumptions usually employed to reduce in-situ data will be tested
against our anisotropic DNS (run B and C) in order to highlight possible systematic errors 
on the measure of the cascade rate in the solar wind.

The simplest assumption is that of isotropic turbulence, Equation~\eqref{PPiso},
whose expression is repeated here for convenience,
\begin{eqnarray}
\epsilon^{iso}=-\frac{3}{4}\frac{Y_\ell}{\ell}, 
\label{ciso}
\end{eqnarray}
and involves projecting $\vect{Y}$ along the direction of increments
($Y_\ell=\vect{Y}\cdot\vect{\ell}/\ell$).
As discussed in section~\ref{sec2}, the cascade
rate is directly obtained from the III-order SF without any need to compute
derivatives. This ``isotropic'' method has been mainly employed for fast polar wind \citep{SorrisoValvo_al_2007,Marino_al_2008,Marino_al_2012} on Ulysses measurements.

Admitting some form of anisotropy, the minimal assumption is that of 
axisymmetry. 
To avoid computing derivatives along the two increments,
one again resorts on simplified
geometrical models, as the 2D+1D turbulence employed by \citet{Macbride_al_2008,Stawarz_al_2009,Stawarz_al_2010} on WIND
and ACE data in the ecliptic solar wind.
This method assumes that the III-order SF has
parallel and perpendicular components that depend only on the parallel and
perpendicular increments, respectively (the cascade is independent in the
two directions). The total cascade rate is obtained by combining the two
independent equations for (isotropic) 2D-perpendicular and for 1D-parallel
cascades (Equation~\eqref{PP2DI} and Equation~\eqref{PP1D} respectively). 
Such a ``hybrid'' cascade rate reads
\begin{eqnarray}
\epsilon^{hybr}=\epsilon^{1D}+\epsilon^{2D}=\epsilon^{||}+\epsilon^\bot=-\left(\frac{1}{4}\frac{Y_{||}}{\ell_{||}}+\frac{1}{2}\frac{Y_{\bot}}{\ell_{\bot}}\right).
\label{caniso}
\end{eqnarray}
The isotropic and hybrid cascades are obtained by first computing the 3D
III-order SF, and then by applying the corresponding average (isotropy or
axisymmetry).
Similarly, the ``true'' cascade rate is obtained from Equation~\eqref{Yaxis}:
\begin{eqnarray}
\epsilon^{true}=-\frac{1}{4}\frac{1}{2\pi}\int\left(\pp{Y_x}{\ell_x}+\pp{Y_y}{\ell_y}+\pp{Y_z}{\ell_z}\right)\d\phi.
\label{ctrue}
\end{eqnarray}

We now compare in Figure~\ref{fig6} the true cascade rate (thick solid line) 
with the isotropic (thin solid line) and the hybrid (thin dashed line) cascade
rate for run B and C (top and bottom panels, respectively).
All cascade rates are normalized by the dissipation rate $\epsilon$ obtained
directly from spectra, Equation~\eqref{diss}.
Each panel represents a cut in the $(\ell_{||},~\ell_\bot)$ plane taken along a
fixed direction that forms an angle $\theta$ with the mean-field direction.

The isotropic cascade rate is strongly angle dependent, independently of the
run considered. It returns increasing cascade rates for increasing angles, as
can be expected since both runs B and C have a dominant perpendicular cascade.
It overestimates the true cascade rate at large angles
($\theta\gtrsim70^o$) while it underestimates the true cascade rate by a factor
$\approx2$ even at $\theta=45^o$. 
The hybrid method instead performs very well: it does not vary with angle
$\theta$ and yields the correct cascade rate at oblique angles $\theta\gtrsim20^o$.

In the nearly parallel direction, both the hybrid and isotropic methods strongly
underestimate the true cascade rate (by a factor $>10$) or they completely fail in yielding any cascade rate.
Indeed, neither of the two methods is able to account for the dependence of
the parallel component of the III-order SF on the perpendicular
increments, $\vect{Y_{||}}(\ell_\bot)$, which is instead fundamental in our
strongly anisotropic runs (e.g. Figure~\ref{fig2}a). Indeed,
the 1D cascade entering the hybrid method is basically negligible at all angles (red dashed line in fig~\ref{fig6}).  
This implies that the cascade is with a good approximation a 2D cascade in both our anisotropic runs.
On the other hand, a linear scaling for the 1D cascade is
found in the solar wind (e.g. \citealt{Macbride_al_2008}), suggesting that the
III-order SF of solar wind turbulence is less anisotropic than that one of runs B and C. 

\section{Summary and discussion}
We studied the anisotropy of MHD turbulence with or without guide field
($B_0$) by means of structure functions (SF) of II and III order,
in three direct numerical simulations of MHD turbulence at moderate and high
resolution (see table~\ref{table1}). We consider isotropic decaying turbulence
(run A), forced strong turbulence with mean-field (run B), and forced weak
turbulence with mean-field (run C).

We used  II-order SF ($S$) to characterize the anisotropy with respect to the
local mean-field and to the global mean-field. 
The anisotropy with respect to the local
mean-field is scale dependent in all runs (with or without mean-field)
and follows the critical balance prediction $\ell_{||}\propto\ell_\bot^{2/3}$.
This confirms previous theoretical and numerical findings
\citep{GS95,Cho_Vishniac_2000,Maron_Goldreich_2001}, in the local frame the anisotropy grows at smaller and smaller scales.
When instead we used as a reference the global mean-field, we found
isotropy for MHD turbulence without mean-field (run A) and strong
two-dimensionalization for turbulence with mean field,  
with run B being more anisotropic than run C (the small-scale aspect ratio is $\approx10$ and $\approx5$ respectively).
Surprisingly, for run C, we found a scale-dependent anisotropy consistent with
critical balance even when $S$ is computed in the frame attached to the mean
field $B_0$, at variance with previous analysis in DNS \citep{Chen_al_2011}
and in solar wind measurements \citep{Tessein_al_2009} which reported
scale-independent anisotropy. A possible explanation is that run C has a strong
magnetic field ($b_{rms}/B_0=1/5$) and so the local and global frames almost coincide. In favor of this
explanation, the anisotropy becomes scale independent at smaller scales.
However,
there are other ``non-standard'' aspects of run C that could be related to
scale-dependent anisotropy in the global frame: the special 3D spectrum that has an isotropic spectral slope but anisotropic energy levels in parallel and perpendicular direction \citep{Grappin_Muller_2010}, the
strong excitation of pseudo Alfv\'en polarizations (see below), and the
isotropic forcing (freezing) that maintains a magnetic excess at large
scales \citep{Grappin_al_2013}. These three aspects could be all related to one another and are under investigation.

We then analyzed the vectorial III-order SF (or Yaglom flux, $\vect{Y}$), which
is related to $S$ computed in the global frame by a generalization to
incompressible MHD \citep{PP} of the Von Karman-Howart-Yaglom relation in
hydrodynamics (KHYPP equation in the following). The III-order SF, $\vect{Y}$, is
expected to be anisotropic for MHD turbulence with mean-field, but it has never
been reported in DNS or experiments (except for rotating hydrodynamic
turbulence experiments, \citealt{Lamriben_al_2011}). Despite some anisotropic models exist \citep{Podesta_al_2007,Galtier_al_2009,Galtier_2012}, they have never been verified in DNS.

We tested the KHYPP equation in isotropic decaying turbulence (run A) and
found that the Yaglom flux is almost purely radial, its divergence is
isotropic, and the maximum of $-\bnabla\cdot\vect{Y}$ is of the order of the
dissipation rate (only a factor 0.8 smaller), confirming theoretical expectations \citep{PP}.
However, the condition defining the inertial range, 
$-\bnabla\cdot\vect{Y}=const$, is only fairly satisfied in our decaying turbulence that has an inertial range of only half a decade despite the high resolution ($1024^3$). 
More interestingly, we showed that the familiar concept of Kolmogorov cascade
in Fourier space also applies in the real space with SF. Note that no
assumption on locality of nonlinear interactions is needed to obtain the KHYPP
equation. Despite this, we found that injection, cascade, and dissipation occur at the same rate and that the cascade is achieved via a fairly constant nonlinear transfer in the inertial range. 
In particular, by comparing the divergence of the Yaglom flux with the
other terms appearing in the KHYPP equation (dissipation term and decaying term), one
can determine the dissipation scale and the injection scale of turbulence.
This has important applications in solar wind
turbulence, since, due to expansion, it is an intrinsically decaying turbulence
for which the injection scale is not easily identified
\citep{Hellinger_al_2013,Dong_al_2014}.

For strong turbulence with mean-field, we found that 
$\vect{Y}$ has only perpendicular components and depends only on
perpendicular increments; this is a 2D turbulence as defined in Equation~\eqref{PP2D}.
The corresponding inertial range extends to small perpendicular scales, while it is shorter and located at larger scales in the parallel direction: the
anisotropy of $\vect{Y}$ emerges as a non-uniform distribution of the scales having $-\bnabla\cdot\vect{Y}=const$.  
A similar non-uniform distribution is found in the forced, weak turbulence
case (run C), with an important difference concerning the orientation of $\vect{Y}$. 
The latter is no more strictly perpendicular, but oblique. In particular,
the Yaglom flux is almost radial at large parallel scales in the inertial
range and becomes more and more perpendicular at small parallel scales. 
We interpret this behavior as a weaker two-dimensionalization of turbulence,
since the oblique orientation of $\vect{Y}$ implies a dependence on parallel
and perpendicular scales, in contrast to the purely 2D case in which
$\vect{Y}=\vect{Y_\bot}(\vect{\ell_\bot})$.

By decomposing fluctuations into shear Alfv\'en and pseudo Alfv\'en polarizations in run C we also
found that energetically important pseudo polarizations are responsible 
for the non-perpendicular III-order SF.
The pseudo/shear decomposition allows us to prove that the two polarizations
cascade independently, that is, without exchanging their energy, in the
inertial range. One can
thus write separate KHYPP equations for pseudo and shear energies. However,
strictly speaking, the two cascades are not independent, since in each of
them the strain of pseudo and shear polarizations enters the expression for
the Yaglom flux. In particular the strain exerted by pseudo polarizations is
associated with a non-constant cascade of energy at the large inertial-range
scales; thus pseudo polarizations control the injection of energy in run C
(although becoming passive at smaller inertial-range
scales, e.g. \citealt{Maron_Goldreich_2001,Cho_Lazarian_2003}). 
In \citet{Grappin_al_2013}, it is shown how the equipartition of kinetic and magnetic energy at the large frozen scales is associated to the weaker
anisotropy of the 3D spectrum of run C, having the property
of a 3D Iroshnikov-Kraichnan turbulence.
Here we argue that kinetic/magnetic equipartition at the largest scales
induces an exchange of energy between pseudo and shear
polarizations, thus allowing pseudo polarizations to control the cascade at the interface between injection and inertial-range scales. 

Finally, on the anisotropic runs B and C we tested two methods that are commonly applied to obtain the cascade rate in the solar wind. 
The methods both rely on assumptions of the
anisotropy of the III-order SF, allowing one to obtain the cascade rate directly
from $\vect{Y}$ (i.e. without computing the divergence).
The first method assumes that $\vect{Y}$ is isotropic.
The second method (hybrid method) assumes axisymmetry and models the 
anisotropy as a geometric superposition of a 1D (parallel) cascade and an
isotropic 2D (perpendicular) cascade.
We found that the isotropic method is strongly angle dependent, yielding correct
cascade rates only at large angles between the direction of increments and the
magnetic field, translating to angles between the magnetic field direction
and the solar wind direction $\theta_{BV}\gtrsim70^o$ (needless to say, the isotropic method works very well for run A).
For fast polar wind, emanating from stable coronal holes, turbulence is expected to have a strong anisotropy
\citep{Verdini_al_2012,Perez_Chandran_2013}, comparable to our run B and
C. We thus suggest that the angle dependence of the isotropic method is at the origin of the relatively small number of intervals ($13\%$) in which the linear
scaling $Y_\ell\propto\ell$ is found in polar solar wind in-situ data
\citep{Marino_al_2012}. 
In contrast, the hybrid method performs very well on our anisotropic
runs, yielding an angle-independent cascade rate for $\theta_{BV}\gtrsim20^o$.
Both methods fails to obtain any cascade rate when increments are
parallel to the mean-field. This is due to the dependence
$\vect{Y_{||}}(\ell_\bot)$ that is excluded in both assumptions but present in all our runs. 
This, together with the good performance of the hybrid method, indicates that
the anisotropic runs B and C are fairly well described by a 2D cascade model, despite their different degree of anisotropy. 
Note that in the solar wind a 1D cascade, actually a linear scaling
$Y_{||}\propto\ell_{||}$, is indeed measured with the
hybrid method, indicating that solar wind turbulence in the ecliptic has a
different and probably much weaker anisotropy compared to our DNS. 

We conclude by noting that the ratio $b_{rms}/B_0\approx1/5$ employed
in our simulation is lower than the actual value found in the solar wind, which
is $\approx 1/2$ at scale of few hours. The small value of fluctuations'
amplitude was chosen to highlight the effects of anisotropy and makes the 
comparison of our results with solar wind turbulence at large scales a bit
weak, being more meaningful for scales shorter than a minute.
In a future work we plan to apply a similar analysis to simulations with
larger ratio $b_{rms}/B_0$ also dropping the assumption
of axisymmetry. In fact, Cluster observations \citep{Narita_al_2010}
showed that the solar wind turbulence is not axis-symmetric (but see
\citealt{Turner_al_2011} for an explanation in term of an observational bias).
DNS of MHD turbulence in the framework of the Expanding Box Model
\citep{GVM93,Grappin_Velli_1996} also show that
expansion causes non axis-symmetric 3D spectra \citep{Dong_al_2014}. Note that
expansion is also responsible for a differential decay of fluctuations, thus
cross helicity and magnetic/kinetic imbalance of fluctuations also enter the
KHYPP equation in the expanding solar wind \citep{Hellinger_al_2013,Gogoberidze_al_2013}.
Another interesting direction is to explore the role of velocity shears
on the KHYPP equation \citep{Wan_al_2009,Wan_al_2010}, which have been invoked to be
the main driver for the turbulence cascade in both the ecliptic
\citep{Roberts_al_1992} and the polar solar wind \citep{Marino_al_2012}. Again,
numerical simulations with the Expanding Box Model seem to be a very promising tool
to understand the combined effect of shear and expansion
(e.g. \citealt{Roberts_Ghosh_1999}) in shaping the cascade of solar wind turbulence.

\textit{Acknowledgments} 
This project has received funding from the European Union's Seventh
Framework Programme for research, technological development and demonstration
under grant agreement no 284515 (SHOCK). Website: project-shock.eu/home/,AV
acknowledges the Interuniversity
Attraction Poles Programme initiated by the Belgian Science Policy Office (IAP
P7/08 CHARM). PH acknowledges grant P209/12/2023 of the Grant Agency of the Czech Republic.
Computational resources were provided by CINECA (grant 2014 HP10CLF0ZB and HP10CNMQX2M).


\end{document}